\documentclass[fleqn,usenatbib]{mnras}

\usepackage{newtxtext,newtxmath}
\usepackage[T1]{fontenc}
\usepackage{ae,aecompl}
\usepackage{epsfig}
\usepackage{color}
\usepackage{rotating}

\usepackage{graphicx}	% Including figure files
\usepackage{amsmath}	% Advanced maths commands
\usepackage{amssymb}	% Extra maths symbols

\newcommand{\cmsq}{\hbox{cm$^{-2}$}}

\newcommand{\lumin}{\hbox{erg~s$^{-1}$}}

\newcommand{\nh}{\hbox{${N}_{\rm H}$}}
\newcommand{\msun}{\hbox{${M}_{\odot}$}}

\newcommand{\be}{\begin{equation}}
\newcommand{\ee}{\end{equation}}
\newcommand{\ba}{\begin{eqnarray}}
\newcommand{\ea}{\end{eqnarray}}

\newcommand{\hst}{\emph{HST}}

\newcommand{\swift}{\emph{Swift}}

\newcommand{\simgt}{\lower 2pt \hbox{$\, \buildrel {\scriptstyle >}\over {\scriptstyle\sim}\,$}}
\newcommand{\simlt}{\lower 2pt \hbox{$\, \buildrel {\scriptstyle <}\over {\scriptstyle\sim}\,$}}
\newcommand{\ls}{\lower 2pt \hbox{$\;\scriptscriptstyle \buildrel<\over\sim\;$}}
\newcommand{\gs}{\lower 2pt \hbox{$\;\scriptscriptstyle \buildrel>\over\sim\;$}}
\newcommand{\sarc}{$^{\prime\prime}\!\!.$}

\newcommand{\mrk}{Mrk\,231}

\title[Swift Monitoring Observations of Mrk 231]{Swift Monitoring Observations of Mrk 231: Detection of Ultraviolet Variability}

\author[K. T. Smith et al.]{
Keith T. Smith,$^{1}$\thanks{E-mail: mn@ras.org.uk (KTS)}
A. N. Other,$^{2}$
Third Author$^{2,3}$
and Fourth Author$^{3}$
\\
$^{1}$Royal Astronomical Society, Burlington House, Piccadilly, London W1J 0BQ, UK\\
$^{2}$Department, Institution, Street Address, City Postal Code, Country\\
$^{3}$Another Department, Different Institution, Street Address, City Postal Code, Country
}

\author[Yang et al.]{Lilan Yang$^{1}$,
Xinyu Dai $^{2}$,
Youjun Lu $^{3,4}$,
Zong-Hong Zhu $^{1,5}$,
Francesco Shankar $^{6}$
\\
$^{1}$School of Physics and Technology, Wuhan University, Wuhan 430072, China; yang\_lilan@mail.bnu.edu.cn; zhuzh@whu.edu.cn  \\
$^{2}$ Homer L. Dodge Department of Physics and Astronomy,University of Oklahoma, Norman OK, 73019, USA; xdai@ou.edu \\
$^{3}$National Astronomical Observatories, Chinese Academy of Sciences, 20A Datun Road, Beijing, 100101, China; luyj@nao.cas.cn \\
$^{4}$School of Astronomy and Space Sciences, University of Chinese Academy of Sciences, No.\ 19A Yuquan Road, Beijing 100049, China\\
$^{5}$ Department of Astronomy, Beijing Normal University, Beijing, 100875, China; zhuzh@bnu.edu.cn \\
$^{6}$ Department of Physics and Astronomy, University of Southampton, Highfield SO17 1BJ, UK; F.Shankar@soton.ac.uk
}

\date{Accepted XXX. Received YYY; in original form ZZZ}

\pubyear{2018}

\begin{document}
\label{firstpage}
\pagerange{\pageref{firstpage}--\pageref{lastpage}}
\maketitle

\begin{abstract}
 We analyze 168 \emph{Swift} monitoring observations of the nearest broad absorption line quasar Mrk\,231 in the UV and X-ray bands, where we detect
 significant variability in the UV ($\sim$2246\AA) light curve with a null probability of $4.3\times10^{-10}$ for a constant model.
 Separately, from an archival sample of \emph{Swift} observed active galactic nuclei (AGN), we measure the relation between UV excess variance and luminosity, finding that the normalized UV excess variance decreases with luminosity.
Comparing to this mean relation, the normalized UV excess variance of Mrk\,231 is smaller, however within the scatter characterising the full population.
The upper limit of the X-ray excess variance is consistent with other AGN.
The power spectrum density of the UV light curve can be well fit by a power law model with a slope of $1.82\pm0.14$ between $10^{-7.5}$ and $10^{-6}$~Hz, consistent with those for typical AGN, with no obvious quasi-periodical oscillation peaks.
 The UV variability and its power spectrum suggest that a significant amount of the UV emission of Mrk\,231 is from the accretion disk.
 The consistencies in the normalized UV variability and the shape of the power spectrum density between Mrk\,231 and other normal AGN suggest that the origin of UV variability of broad absorption line quasars is similar to other AGN, and dust scattering at large scales such as the torus is not a dominating process for the UV emission of Mrk\,231.  Significant scattering, if present, is constrained to smaller than $\sim$10 light days.
We perform lagged correlation analysis between the UV and X-ray light curves and find the correlation insignificant within the present data.
\end{abstract}


%%%%%%%%%%%%%%%%%%%%%%%%%%%%%%%%%%%%%%%%%%%%%%%%%%

%%%%%%%%%%%%%%%%% BODY OF PAPER %%%%%%%%%%%%%%%%%%

\section{Introduction} \label{sec:1}

Timing analysis is a powerful tool to study the properties of active galactic nuclei (AGN), since they in general exhibit variability on a vast range of time scales 
and wavelengths.
In situations when the data is sparse, the noise removed variance of the light curve is often used as a first step to characterize the variability \citep[e.g.,][]{Nandra1997, Vaughan2003}.   In general, it is found that luminous AGN have relatively smaller variability amplitudes compared to the less luminous ones \citep[e.g.,][]{Nandra1997,1998ApJ...501L..37P,2000MNRAS.315..325A,2002MNRAS.330..390M,2004ApJ...605...45D}.
  When the data quality is high and densely sampled, Fourier analysis can be applied to obtain the power spectrum density (PSD) of the light curve.  Previous studies found that the PSD of AGN is generally composed of a single or multiple segments of power laws \citep[e.g.,][]{2002MNRAS.332..231U,2003ApJ...593...96M,2011ApJ...743L..12M,2013MNRAS.430L..49M,2014ApJ...795....2E}.
There are also situations in between, and the data is used to test a particular PSD model, e.g., the damped random walk model, which has a PSD power law slope of $-2$ but flattening at low frequencies.
Many studies have concluded that the optical AGN variability is consistent with the damped random walk model \citep[e.g.,][]{2009ApJ...698..895K, 2010ApJ...721.1014M, 2010ApJ...708..927K}.
Most previous PSD analyses of AGN focus on the visible or X-ray light curves, and in very few cases densely-sampled UV light curves are available for the PSD analysis.  This is especially true for the sample of broad absorption quasars, which show broad, blue-shifted, absorption troughs in their spectra, and it is estimated that broad absorption line quasars contribute to 20--40\% of Type I quasar population intrinsically \citep[e.g.,][]{2008ApJ...672..108D, 2008ApJ...687..859S,2011MNRAS.410..860A}.  Broad absorption line quasars are also observed to be either X-ray weak or intrinsically X-ray weak \citep[e.g.,][]{1996ApJ...462..637G,2000ApJ...533L..79M,2002ApJ...567...37G,2003AJ....126.1159G, 2011ApJ...737...46M, 2014ApJ...786...58M}.

Correlations between multi-band continuum light curves can reveal the origin of the disk variability.
As reviewed by, e.g., \citet{MH16}, the origin of the UV-optical variability in AGN is still a matter of debate. Leading theories broadly suggest either a reprocessing of X-ray emission by the accretion disc, or simply intrinsic variability of the thermal emission by the disc itself. Reverberation mapping has been adopted as a valid discriminator between these two proposals. For example, evidence of a time lag between the UV/optical variations occurring later than the X-ray ones would be in support of the former scenario.

The ultra-luminous infrared galaxy Mrk\,231 hosts the nearest quasar at a redshift of $z=0.042$, where
the AGN contribution of the bolometric luminosity is estimated to be $10^{46}$~\lumin\,\citep[e.g.,][]{2009ApJS..182..628V}.
Mrk\,231 is also a broad absorption line quasar with low-ionization lines including Fe absorption troughs \citep{1972ApJ...173L.109A}, and thus is classified as an FeLoBAL. FeLoBALs contribute to 2\% of Type I quasar population intrinsically \citep{2012ApJ...757..180D}.
Mrk\,231 is intrinsically X-ray weak as revealed by \emph{Nustar} observations, the X-ray absorption column density is constrained to be heavily absorbed but Compton-thin ($\nh = [1.2\pm0.3] \times 10^{23}~\cmsq$), and the intrinsic X-ray spectrum is flat with a power-law photon index of $\Gamma=1.4$ \citep{2014ApJ...785...19T}.

Overall, Mrk\,231 exhibits many unusual features among quasars.
Recently, an additional unusual
spectral feature was discovered in the ultraviolet (UV) bands
\citep{2013ApJ...764...15V,2014ApJ...788..123L,2015ApJ...809..117Y}.
The continuum of Mrk\,231 drops dramatically from the optical to the
near ultraviolet (NUV) band; however, the drop stops at the
far ultraviolet (FUV) band, and FUV emission is detected all the way to $\sim$1000\AA.
There are several models to interpret the anomalous UV spectrum of \mrk.
It is potentially possible to distinguish these models from the UV variability of \mrk, in addition to measure the overall UV variability characteristics of a broad absorption line quasar.
\citet{2015ApJ...809..117Y} measured
UV emission of Mrk\,231 by
analyzing the archival International
Ultraviolet Explorer (\emph{IUE}) and Hubble Space Telescope (\emph{HST}) data, and found it to be variable.
\citet{2016ApJ...829....4L} and
\citet{2016ApJ...825...42V}, however, argued that the UV emission of
Mrk\,231 may not vary based on analyses on \emph{HST}
observations only.
It is therefore important to check whether
the UV emission from Mrk\,231 varies or not.

Most of the limited UV observations of Mrk 231 are spectroscopic.
These include three spectroscopic measurements by \emph{IUE} between 1978--1980 and a few spectroscopic
measurements by \emph{HST} from 1996 to 2014. Recently, the Neil Gehrels Swift Observatory (hereafter \swift) has made a large number of imaging UV
observations on Mrk\,231 during the period from 2012 to 2016, providing
accurate photometric flux measurements. In this paper, we analyze the
\emph{Swift} UVOT and X-ray data to study the variability of
Mrk\,231 in these bands.
In Section\,\ref{sec:2}, we describe the method to process \emph{Swift} data.
We analyze the light curves obtained from \emph{Swift} UV and X-ray data in Section\,\ref{sec:3}.
Conclusions and discussion are given in Section\,\ref{sec:4}.

\section{\emph{Swift}  Data} \label{sec:2}
\emph{Swift} performed an intense monitoring campaign on Mrk\,231 during 2016 with a total of 175 observations, where two observations do not have UVOT exposures and five observations have pointing errors.
In addition, there are six sparse \emph{Swift} observations of Mrk\,231 between 2012 to 2015.
Since several of these six earlier observations are affected by the ``dropout'' problem in the UV image, we focus on the 2016 data only in this paper with 168 valid observations.

The ultraviolet/optical telescope \citep[UVOT,][]{2005SSRv..120...95R} onboard \emph{Swift} \citep{2004ApJ...611.1005G} has six filters, i.e., uvw2, uvm2, uvw1, u, b, and v, with effective
wavelengths at 1928\AA, 2246\AA, 2600\AA, 3465\AA, 4392\AA,  and
5468\AA, respectively. In the 2016 monitoring campaign, the narrower UV filter, uvm2, which is not affected by the ``red-leak'' issue \citep[e.g.,][]{2010ApJ...721.1608B}, is used in most of the monitoring observations.  We used the software tools included in the UVOT package (Heasoft Version 6.18) to analyze the uvm2 images.
For each observation, the tool \textit{uvotimsum} was used first to sum images and exposure maps.
The source region of Mrk\,231 core is set in a circle with a
radius of 5\arcsec \citep[following][]{2015AJ....149...85G}.
An annulus region with inner and
outer radii of (40\arcsec, 60\arcsec) was
selected as the background region.
UVOT magnitudes and fluxes were measured with the task
\textit{uvotsource} based on the most recent UVOT calibration as described in \citet{2008MNRAS.383..627P,2010MNRAS.406.1687B}.
We corrected the coincidence lost and large-scale sensitivity effects for these measurements, but did not perform an aperture correction since we are interested in the variability of the source in this paper rather than the absolute flux.  For the same reason, we did not include systematic uncertainties associated with the zero-point calibration; turning the \textit{syserr} argument on would artificially increase the uncertainties of the light curve between data points as mentioned by the \emph{Swift} UVOT Software Guide.

Although the UVOT magnitudes have absolute flux calibrations with zero-point uncertainties from 0.03 to 0.01~mag from uvw2 to uvv bands \citep{2011AIPC.1358..373B},
the \emph{Swift} UV photometry can be affected by a ``dropout'' problem.  When the source is located at particular regions on the detector, the measured flux can be underestimated by 0.1~mag or more \citep{2015ApJ...806..129E}.
We removed the potential ``dropouts'' using the following criteria:
if $(m_{N+1}+m_{N-1})/2-m_{N} < -0.07$\,mag, where $m_{N}$ is the uvm2 magnitude of Mrk\,231 core of the ${N_{th}}$ visit, $m_N$ is considered to be a potential ``dropout'' and was discarded in the following analysis.
In the end, we have 127 data points left after these screenings, and we list the uvm2 magnitudes
in Table~\ref{tab:obs}.

We reprocessed the X-ray Telescope \citep[XRT,][]{2005SSRv..120..165B} images of Mrk\,231 taken in the photon counting mode, using \textit{xrtpipeline} version 0.13.2.
The exposure maps were calculated with the tool \textit{xrtexpomap}.
The source and background count rates were extracted using \textit{XSELECT} version 2.4c, 
source region of Mrk\,231 was selected to be a circle with
a radius of 24\sarc8 centered on the X-ray source \citep[following][]{2015AJ....149...85G}, and the background region was selected from a nearby circular region with a radius of 99\sarc2.
Gehrels statistics \citep{1986ApJ...303..336G} was used to calculate the uncertainties.
We list the background subtracted count rates of Mrk\,231 in 0.3--8~keV band in Table~\ref{tab:obs} as well.
We did not further separate the X-ray data into a soft and hard band because of the poor signal-to-noise ratio.

\section{Variability, Power Spectrum, and Correlation Analysis on the UV and X-ray Light Curves }
\label{sec:3}

\begin{figure}
\includegraphics[height=0.39\textwidth]{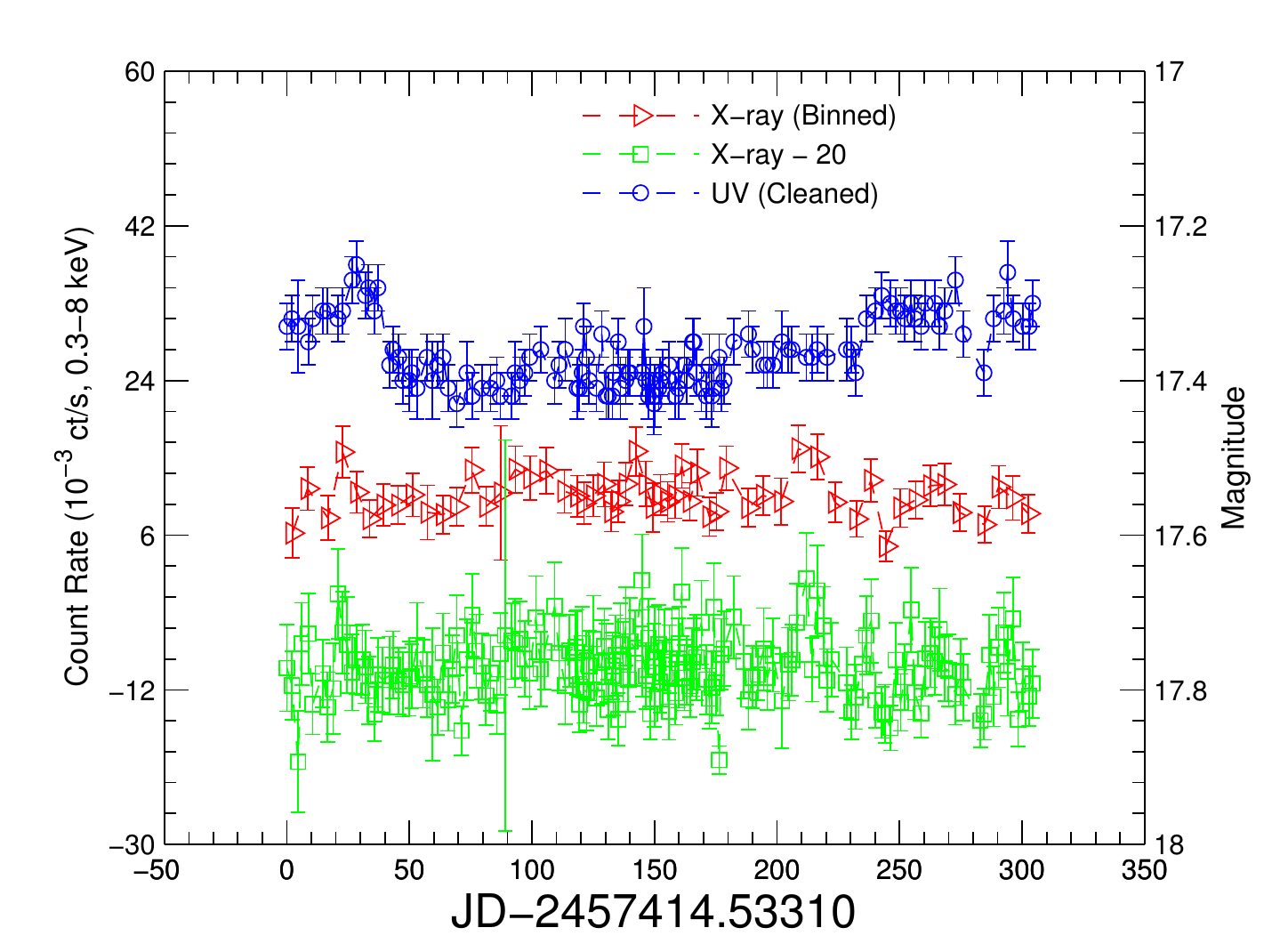}
\caption{Light curves of Mrk\,231 core in uvm2 filter (top), X-ray band (0.3--8~keV) with a bin size 3 data points (middle), and the unbinned X-ray light curve (bottom).  The UV light curve is significantly variable based on the $\chi^2$ test result.}
\label{fig1}
\end{figure}

\begin{figure*}
 % \epsscale{1.17}
%  \includegraphics[height=0.48\textwidth]{L_sigma_uv_2.eps} \includegraphics[height=0.48\textwidth]{L_sigma_2.eps}
  \includegraphics[height=0.48\textwidth]{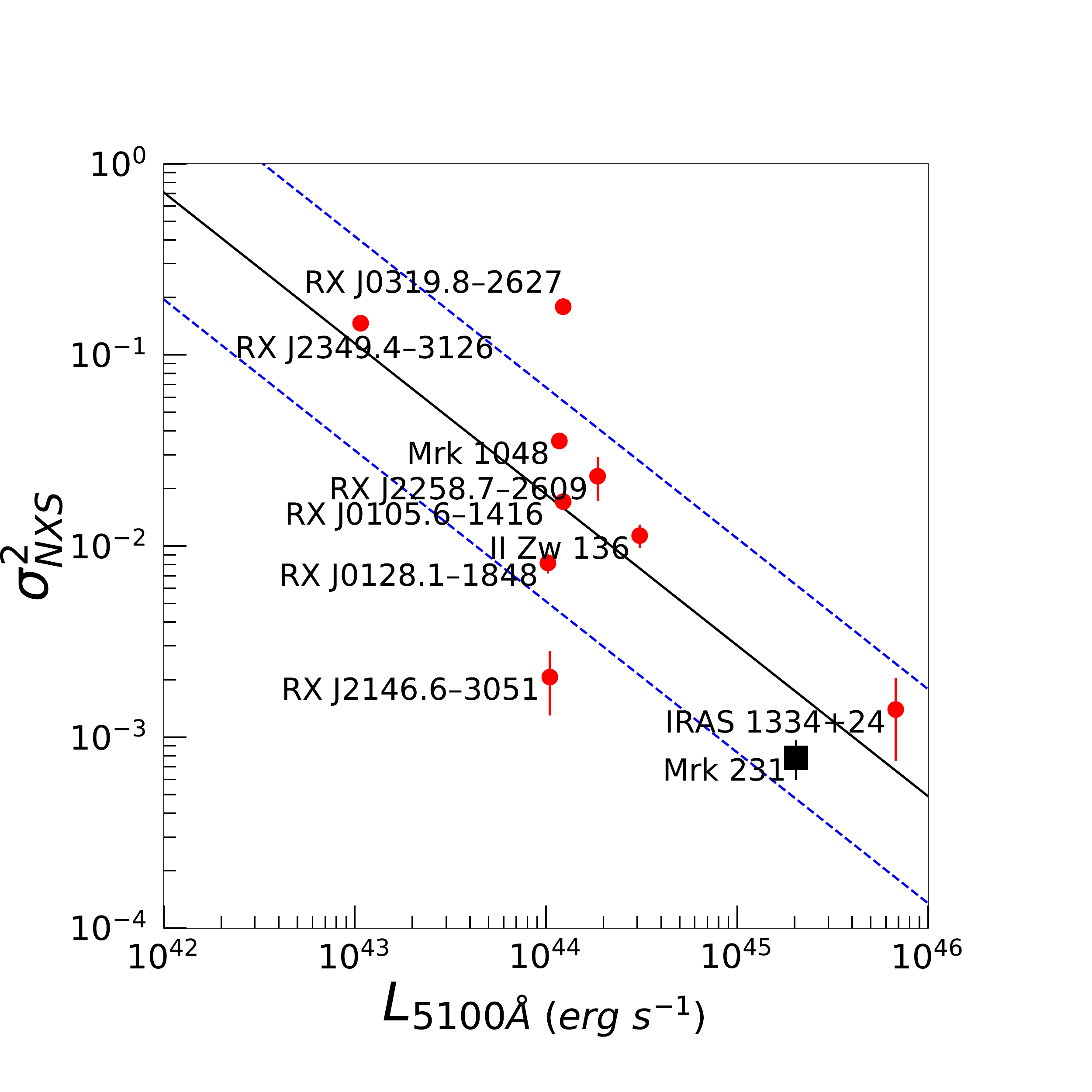} \includegraphics[height=0.48\textwidth]{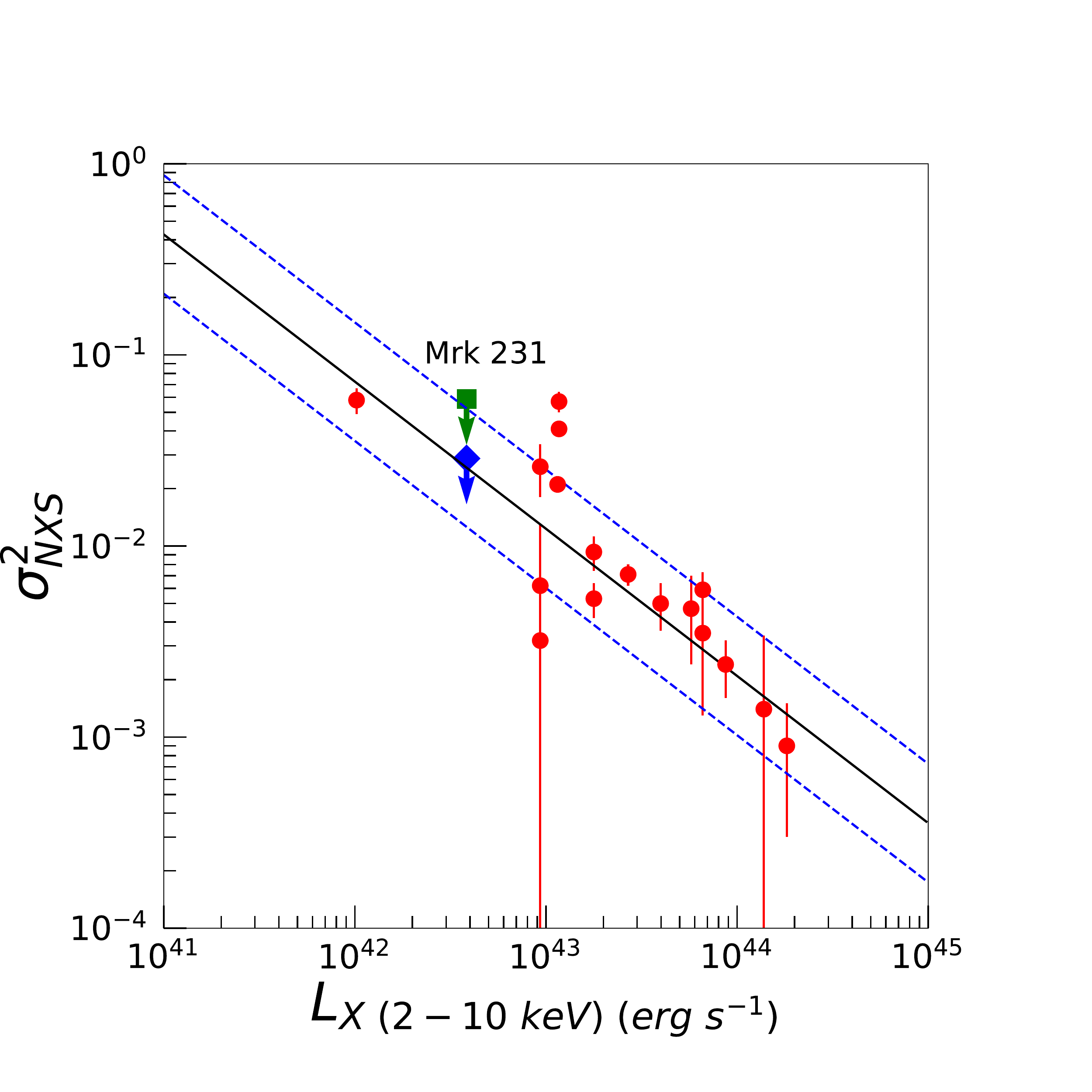}
 % \plottwo{L_sigma_uv_2.eps}{L_sigma_2.eps}
  \caption{Left: UV normalized excess variance versus 5100\AA\ luminosity for Mrk\,231 and a sample of AGN observed by \swift.  Right: X-ray normalized excess variance versus 2--10~keV luminosity. The blue diamond and the green square represent the $1\sigma$ and  $3\sigma$  upper limits of Mrk\,231 respectively. The solid and dashed lines show the mean relation and intrinsic scatter range in both plots.}
\label{fig2}
\end{figure*}

\begin{figure}
\includegraphics[height=0.48\textwidth]{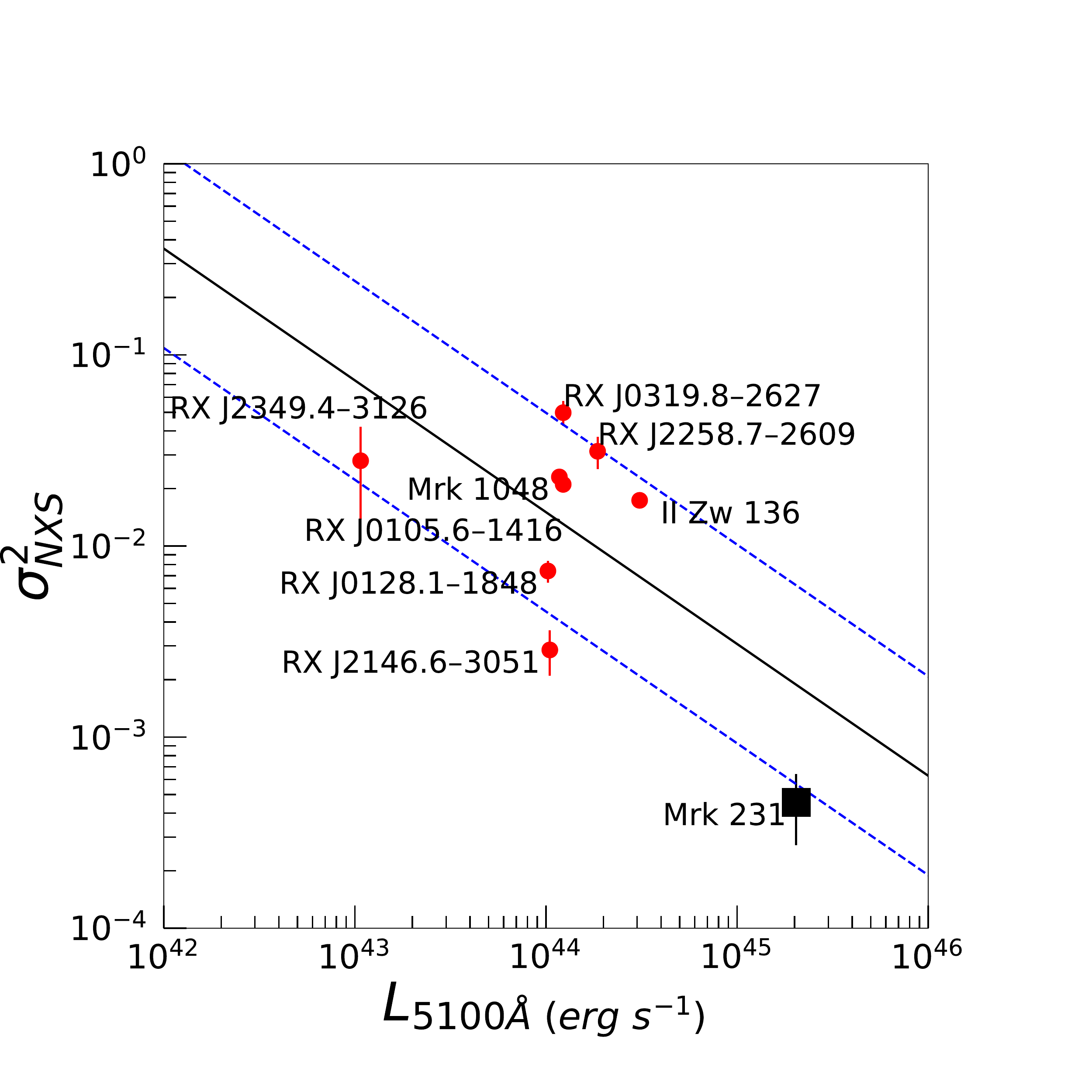}
\caption{ The corrected UV normalized excess variance versus 5100\AA\ luminosity for Mrk\,231 and a sample of AGN observed by \swift.}
\label{fig2_corrected}
\end{figure}

Figure~\ref{fig1} shows the UV (uvm2 filter centered at 2246\AA), X-ray (0.3--8~keV), and binned X-ray light curves of Mrk\,231 core.
 We first applied a $\chi^2$-test to check the variability of Mrk\,231 core both in the UV and X-ray bands.
The UV light curve shows significant variability with a $\chi^2$ null probability for a constant model of $4.3\times 10^{-10}$.  The unbinned X-ray light curve does not show significant variability with a null probability of 0.59.
Binning the X-ray light curve by bin sizes of 3--7 data points, the $\chi^2$ null probabilities are in the range of 0.6--0.9.  Thus, the X-ray light curve does not show significant variability from the current data.
To quantify the UV variability, we calculated the normalized excess variance based on \citet{Vaughan2003}:
\begin{equation}
\sigma^2_{NXS}=\frac{S^2-\overline{\sigma^2_{err}}}{\bar{x}^2},
\end{equation}
with its uncertainty
\begin{equation}
err(\sigma^2_{NXS})=\sqrt{\left(\sqrt{\frac{2}{N}}\cdot\frac{\overline{\sigma^2_{err}}}{\bar{x}^2}\right)^2+\left(\sqrt{\frac{\overline{\sigma^2_{err}}}{N}}\cdot\frac{2F_{var}}{\bar{x}}\right)^2},
\end{equation}
where $S^2$ is the sample variance, $\overline{\sigma^2_{err}}$ is the mean square error and $F_{var}$ is defined as $\sqrt{(S^2-\overline{\sigma^2_{err}})/\bar{x}^2}$.
In the UV band, the normalized excess variance is measured to be $\sigma^2_{NXS}=7.8\times10^{-4}$ with an uncertainty $err(\sigma^2_{NXS})=1.8\times10^{-4}$.
In the X-ray band, we estimated the 3$\sigma$ upper limit to be $5.9\times10^{-2}$.

To compare with UV excess variances of other AGN, we used the published \swift-UVOT light curves of local Seyferts \citep{2010ApJS..187...64G} and calculated the excess variance for those objects with more than \textbf{2} \emph{Swift} observations, excluding narrow line Seyfert 1s (Figure~\ref{fig2}). 
For the X-ray band, we compared our limit with the data from \citet{Nandra1997}.
The UV normalized excess variance decreases with increasing optical luminosities, similar to the trend observed in the X-ray band.  We fit a linear relationship between the UV excess variance and 5100\AA\ luminosity as,
\begin{small}
\begin{equation}
\log{\sigma^2_{NXS, UV}} =(-0.79\pm0.25) \log{\frac{L_{5100\mbox{\AA}}}{\rm erg~s^{-1}}} + (33.03\pm11.02) \pm(0.56\pm0.24),
\end{equation}
\end{small}
where the last term is the intrinsic scatter of the relation.
The UV excess variance of Mrk\,231 is below the mean of the relation, but within the intrinsic scatter range.
We also fit the X-ray data as,
\begin{small}
\begin{equation}
\log{\sigma^2_{NXS, X}} = (-0.76\pm0.15) \log{\frac{L_{X}}{\rm erg~s^{-1}}} + (31.15\pm6.49) \pm(0.31\pm0.09),
\end{equation}
\end{small}
and the X-ray excess variance limit of Mrk\,231 is consistent with the mean relation.

{\cite{2013ApJ...771....9A} have shown that there are biases for measuring the normalized excess variance from both the sampling method and length.
We first corrected the bias due to the sample method, assuming the slope of PSD $\beta=2$ for all AGNs here.
% (see the definition of $\beta$ in formula \ref{bfactor}).
The correction factors for Mrk\,231 and for the ramaining objects are $0.49$ and $0.57$ respectively (see Table 2 in \cite{2013ApJ...771....9A} ).
Next, for the bias from the sampling length, the correction factor \textit{b} is used to 
calculate the intrinsic normalized excess variance. 
The \textit{b} factor is equal to
\begin{equation}
b\propto\left(\frac{T_{max}}{T^\prime_{max}}\right)^{\beta-1} .
\label{bfactor}
\end{equation}
$T_{max}$  in Equation \ref{bfactor}, which we set to $T_{max}=1$ year, is the time assumed reference timescale within which one 
could measure the intrinsic normalized excess variance.
%supposed can measure intrinsic normalized excess variance and we set $T_{max}=1 year$, 
 ${T^\prime_{max}}$ is instead the observation duration.
 The exact value of $T_{max}$ is not important here, since setting a different value will only introduce a constant offset for the whole sample.
%,  and $\beta$ is the slope of PSD. 
To correct the bias, we only consider objects with observations spanning longer than 100 days to avoid large corrections.
After multiplying the correction factor \textit{b}, the  unbiased results are shown in Figure~\ref{fig2_corrected}.
We also fit a linear relationship between the corrected UV excess variance and 5100\AA\ luminosity as,
\begin{small}
\begin{equation}
\log{\sigma^2_{NXS, UV}} =(-0.69\pm0.33) \log{\frac{L_{5100\mbox{\AA}}}{\rm erg~s^{-1}}} + (28.52\pm14.44) \pm(0.52\pm0.25),
\end{equation}
\end{small}
The fitting results are consistent with those without the corrections, and Mrk\,231 is still consistent with the mean relation with scatter.}

\begin{figure*}
%          \fig{psd_uv_ev.eps}{0.45\textwidth}{}
%  \epsscale{1.17}
\includegraphics[height=0.39\textwidth]{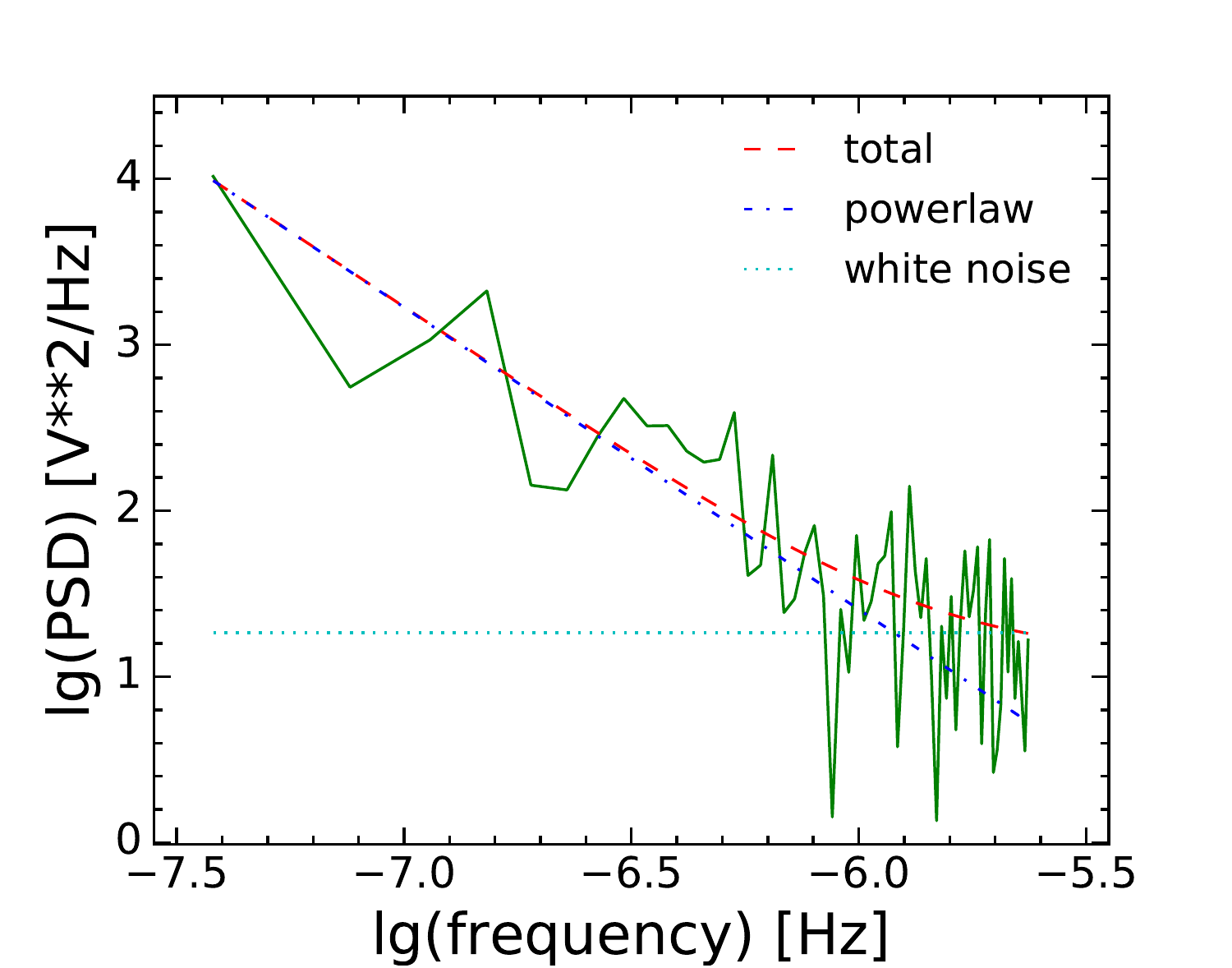}  \includegraphics[height=0.36\textwidth]{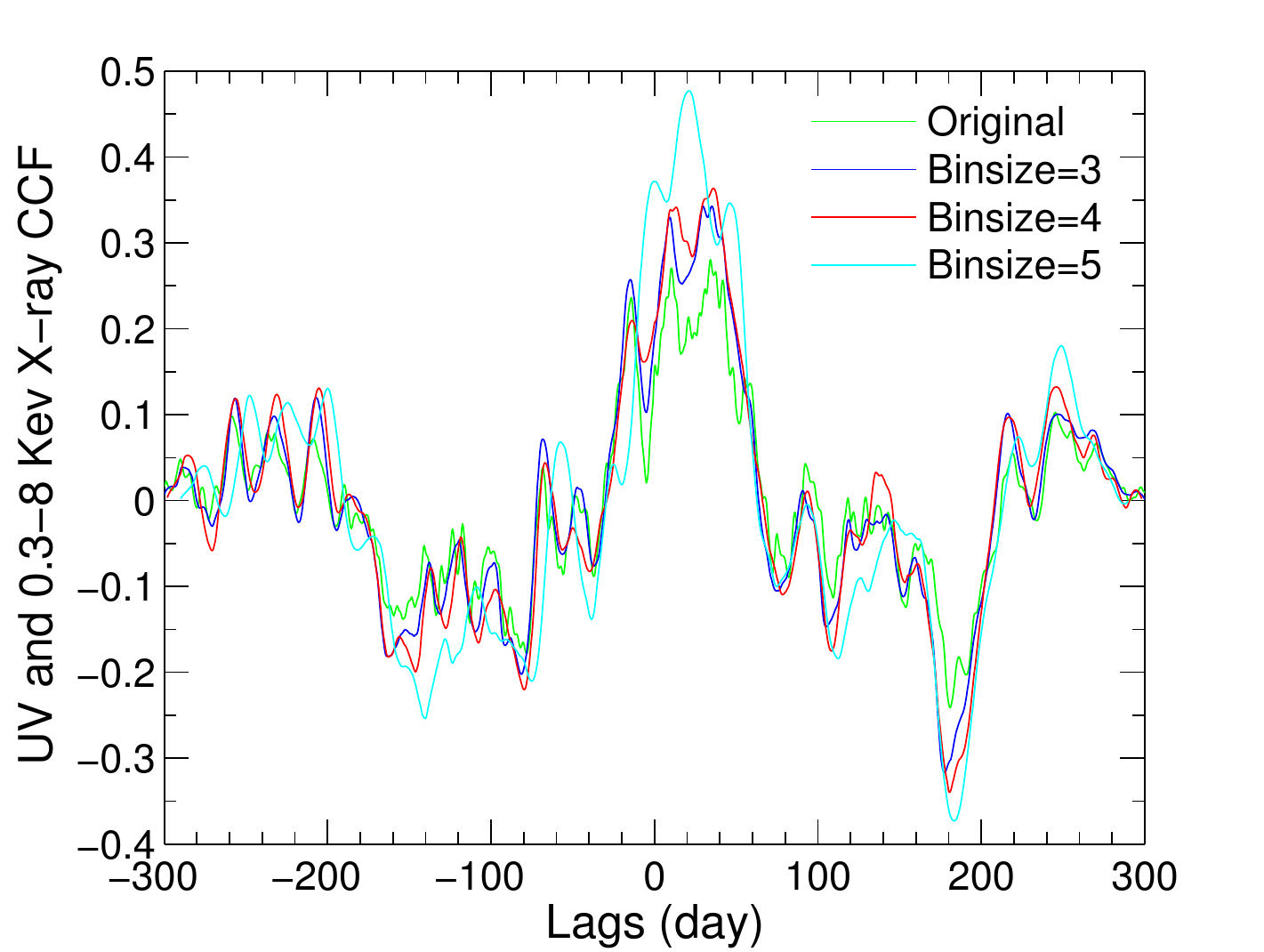}
\caption{Left: the power spectrum density (green solid line) of the UV light curve of Mrk\,231 core fit
  by a power law plus constant noise model (red dashed line).  The fit yields a power law slope of $1.82\pm0.14$, and the blue dash-dotted and cyan dotted lines show the power law and constant components and they intersect at $10^{-5.9}$~Hz.  Right: Lagged cross correlation functions between UV and X-ray light curves of different bin sizes.  The UV and X-ray light curve correlations are insignificant based on simulations.}
\label{fig3}
\end{figure*}

We next analyzed the PSD of the UV light curves (Figure~\ref{fig3}), finding that it can be fit by a power law model $P \propto f^{-\beta}$ with a slope of $\beta = 1.55\pm0.15$.  Since the measurement uncertainties can contribute a white noise component and we can visually see the high frequency flattening above $10^{-6}$~Hz, we performed another fit to frequencies below $10^{-6}$~Hz and obtained a PSD slope $1.8 \pm 0.3$.
Next, we added the constant noise to the fit and yielded a power law slope of $1.82\pm0.14$, and the power law and constant noise components cross at $10^{-5.9}$~Hz.

We finally performed a lagged-correlation analysis between the UV and X-ray light curves.
Although the X-ray light curve does not show significant variability due to its poor S/N, combining the UV and X-ray data might reveal weak signals from the X-ray light curve.
Since the data were not evenly sampled, we performed linear interpolations between the UV and X-ray data points with a step size of 0.5 day to evenly sample the light curves.
After these interpolations, 609 points in each curve are used to calculate the cross correlation functions (CCFs).
The correlation analysis was performed both between the UV and unbinned X-ray light curves and UV and binned X-ray light curves with different bin sizes (the right panel of Figure~\ref{fig3}). 
%\textcolor{red}{We find the 
%, where positive lags indicate that the UV light curve is leading} 
%\textcolor{red}{ARE WE SURE?? OTHERS FIND JUST THE OPPOSITE...}.

The correlation strength between UV and X-ray light curves are $r_{max} = 0.23$ using the unbinned X-ray light curve.  Binning the X-ray light curve by bin sizes of 3, 4, and 5 data points, the corresponding $r_{max}$ values are 0.26, 0.31, and 0.48, respectively.
We evaluated the significance of the correlations by simulating the X-ray light curves with two models, first a white noise only PSD, second a powerlaw PSD with a slope of $1.3$ and with the additional white noise, using the technique of \citet{tk95}.  The normalizations of the model PSDs were set such that the expected values of the total variance match the observational value.  We simulated 168 data points, the same as the observation, and 1000 X-ray light curves in each model and performed lagged-correlations with the observed UV data, and recorded the simulated $r_{max}$ values.  For the white noise null model, 34 cases out of the 1000 simulations have the simulated $r_{max}$ values larger than the observational one.  Binning the simulated data by 3, 4, 5 data points, the numbers that exceed the observations are 75, 253, and 113, respectively.  For the powerlaw PSD with the white noise model, the corresponding numbers are 719, 461, 669, and 387 for the unbinned, and binned simulated light curves.  The simulations show that the UV-X-ray correlation is only marginally significant for the case where the X-ray null PSD model is white noise and the simulated data are unbinned.  In all other cases, the simulations show that the UV-X-ray correlation is not significant.
We conclude that the UV-X-ray correlation is insignificant in the current \emph{Swift} data set for \mrk. 

\section{Discussion and Conclusions}
\label{sec:4}

In this paper, we have analyzed the UV and X-ray observations of Mrk\,231 core by
\emph{Swift} and found that its UV emission
 varies significantly based on the $\chi^2$ test result.
The detection of UV variability for Mrk\,231
is consistent with our previous analysis in the FUV band ($\sim$1030\AA) based on the archival \emph{IUE} and \emph{HST} data
\citep{2015ApJ...809..117Y}, and in contrast with other analyses claiming that the UV flux of Mrk\,231 is a constant.
UV variability from the accretion disk is a characteristic of many AGN observed by \emph{Swift} \citep[e.g.,][]{2010ApJS..187...64G, 2012ApJS..201...10W, 2013A&A...550A..71V}.

The PSD of the UV light curve of Mrk\,231 is well fit by a power law model, another characteristic of AGN variability, where the slope is measured to be $1.8\pm0.3$ for frequencies below the flattening caused by white noise.
Very few AGN have the power spectrum measured in the UV band, and thus we compare with those measured in the optical and X-ray bands.
In the frequency range between $10^{-5.5}$ to $10^{-7.5}$\,Hz, the typical slope for the PSD of AGN is $\beta \sim 1-2$ in the X-ray band \citep[e.g.,][]{2002MNRAS.332..231U, 2003ApJ...593...96M, 2013MNRAS.430L..49M} and $\beta \sim 2-3$ in the optical band \citep[e.g.,][]{2009ApJ...698..895K, 2011ApJ...743L..12M, 2014ApJ...795....2E}.  The UV PSD slope measured in Mrk\,231 is consistent with the damped 
%\textcolor{red}{(we say "damped" but, as far as I can tell, we cannot probe the region in which the PSD in AGN has been shown, such as in SDSS, to be damped wrt an extrapolated pure random walk...)} 
random walk model of AGN optical variability \citep[e.g.,][]{2009ApJ...698..895K, 2010ApJ...721.1014M, 2010ApJ...708..927K}.
Since Mrk\,231 is a broad absorption line quasar, an FeLoBAL in particular, dust is expected to be present in the central engine.  However, the amount of dust, its location, and how it affects the UV emission are all uncertain.  If dust scattering is important for the UV emission of \mrk, we expect that the PSD will be significantly suppressed at frequencies higher than that corresponding to the dust scattering scale.
For broad line regions, the size scale is $\sim$60 light days for $10^8\msun$ black holes \citep[e.g.,][]{2011A&A...528A.100S}, and
for the dusty torus, typical estimates are $\sim$300 light days \citep[e.g.,][]{2006ApJ...639...46S} for quasars.
These scales correspond to light-crossing times with frequencies between $4\times10^{-8}$ and $2\times10^{-7}$~Hz, and we do not measure significant suppression of variability power above these frequencies.  Instead, the PSD extends to $10^{-6}$~Hz following the typical power-law model for AGN.  This suggests that dust scattering at large scales is not a dominating process for the UV emission of \mrk.
The high frequency PSD measurements allow us to constrain the UV emission or scattering scales smaller than $\sim$10 light days.

The  upper limit of the X-ray variability measured for \mrk, with a massive black hole of $\sim 1.5\times 10^8 \msun$
% \textcolor{red}{\citep{2015ApJ...809..117Y, 2005ApJ630122G}}, 
based either on our $H_\alpha$ line width measurement from the optical spectrum reported in \cite{2014ApJ...788..123L} and the virial mass estimator of \cite{2005ApJ630122G} or the BBH fit of \cite{2015ApJ...809..117Y},
is consistent with the general finding that (short) time-scale variability is inversely proportional to black hole mass \citep[][and references therein]{2001MNRAS.324..653L,2013MNRAS.430L..49M}. 
%\textbf{\textcolor{red}{The relationship between $\sigma^2_{NXS}$ and black hole mass was first found in \cite{2001MNRAS.324..653L}.}}
Our findings pointing to an anti-correlation between amplitude of UV variability and optical luminosity is also consistent with the idea that variability is anti-correlated to mass accretion rate,
though most probably additional physical drivers can complicate this basic picture \citep[e.g.,][and references therein]{Ive14}.

The time lag between X-ray and UV-optical fluxes, with the latter usually lagging behind, is broadly consistent with the idea of reprocessing of higher energies photons by an accretion disk. At longer wavelengths, the lag $\tau$ has been shown to increase steadily with increasing wavelength as $\tau\propto \lambda^{\beta}$, with $\beta \gtrsim 1$, as expected from an extended and/or non-uniform disc \citep[][and references therein]{MH17}. Extrapolations of lags to shorter wavelengths/higher energies, does not always agree with the measured X-ray to UV lag \citep{MH17}. \citet{MH14} find long-term UV/optical variations are not necessarily paralleled in the X-rays, suggesting an additional component to the UV/optical variability possibly arising from accretion rate perturbations. \citet{Pal18} studied the variability in the Seyfert 1 galaxy NGC 4593, finding evidence for both a highly variable component such as hard X-ray emission, and a slowly varying disc-like component. Their data suggest the observed variation in longer wavelengths to be due to X-ray reprocessing. \citet{Cac18} monitored the Seyfert 1 galaxy NGC 4593 broadly confirming a lag spectrum $\tau\propto \lambda^{4/3}$ relation consistent with the standard thin disk model. However, they also suggest that larger disk sizes and emission from the Broad Line Region should also be considered as essential components to properly model AGN lag spectra.
We do not detect significant lag between the UV and X-ray light curves from \emph{Swift} observations of Mrk\,231 mostly because of the low signal-to-noise ratio of the X-ray light curve.  A similar monitoring campaign with XMM-Newton will be much more promising to detect the UV and X-ray lag in \mrk, since the UVOT instruments onboard of \emph{Swift} and XMM-Newton are similar while the throughput of X-ray mirror of XMM-Newton is an order of magnitude higher.

We measured the normalized excess variance of Mrk\,231 in the UV band and compared with the values from a sample of nearby AGN observed by \swift.  We performed a linear fit to the normalized excess variance measured in the UV band with the 5100\AA\ luminosity including intrinsic scatter, and we found that Mrk\,231 is below the mean relation from $0.3$ to $0.6$\,dex, but still consistent with other AGN considering the intrinsic scatter of the sample ($\sim 0.6$\,dex).
Although within the scatter, we discuss several other factors that could contribute to \mrk's relatively low variability amplitude compared to the mean relation.

First, Mrk\,231 is a broad absorption line quasar, and the presence of the disk wind may suppress the observed optical/UV variability from the disk. \citet{Conno14} suggest that the complex X-ray variability observed in the Seyfert 1.8 galaxy NGC 1365, could also be interpreted with a wind model in which the launch radius moves out with increasing X-ray luminosity. 
%For BALQSOs, variability studies have mainly focused on the broad absorption line variability usually interpreted as due to changes in photoionization \citep[e.g.,][]{Stern17}. \citet{Wildy14} studied 50 BALQSOs in the redshift range 1.9 < z < 4.2. They did not find signs of correlation between BAL line variability and quasar luminosity.
In addition, it is possible that Mrk\,231 is accreting in the super-soft state as proposed by \citet{2014ApJ...785...19T}, which predicts a low variability amplitude and a very hard X-ray spectrum as well. Indeed, \citet{Conno16} from the X-ray spectral variability of 24 local AGN from the Palomar sample of nearby galaxies, found that AGN with low accretion rates show hardening with increasing count rate, converse to the softer-when-brighter behaviour normally observed in AGN with higher accretion rates. 
%In addition, as reviewed by \citet{MH10}, the rms X-ray variability scales broadly linearly with flux in AGN.
Finally, the contribution of large-scale UV emission can be significant and dilute the variability from the central engine, as indicated by the FUV \hst\ image that Mrk\,231 is slightly extended \citep{2016ApJ...829....4L}.  If the large-scale UV emission contributes to 30-50\% of the observed UV luminosity, the normalized excess variance of Mrk\,231 will be on the mean relation.

As recently reviewed by \citet{P17}, high column densities of circumnuclear absorbing material can also significantly affect the fraction and extension of the X-ray variability in AGN. \citet{Gonza18} found non-trivial variations in 19 out of 22 X-ray-selected AGN, concluding that obscuration along the line of sight is an important parameter in shaping the observed correlations between, e.g., black hole mass, accretion rate, and break frequencies. \citet{H15} find from a sample of 26 Seyfert 2 galaxies that short-term X-ray variability is mostly associated to Compton-thin sources, which should more safely arise from variations in the nuclear source. In their sample UV variability on longer timescales seems not to be affected by the level of line-of-sight obscuration. \citet{Sanchez17} presented variability for a large sample of X-ray-selected AGN in COSMOS with different levels of obscuration. They found that broad-line AGN have a larger fraction of variable sources than narrow-line ones, and that X-ray-classified unobscured AGN tend to have a lower fraction of variable sources with respect to optically-classified unobscured AGN, possibly due to differences in the origin of the obscuration.

Combining the variability, PSD, and excess variance analysis, we conclude that  a significant fraction of the observed UV emission of Mrk\,231 in uvm2 band centered at 2246\AA\ is from the accretion disk.
These dense \emph{Swift} UV monitoring data of Mrk\,231 provide the first opportunity to characterize the UV variability from a broad absorption line quasar in detail, although Mrk\,231 belongs to the rarer sample with Fe absorption troughs.  We found that its fractional variability and shape of the PSD are consistent with other normal AGN.  This suggests that the origin of UV variability of broad absorption line quasars is similar to other AGN.

The central wavelength is shorter than the sharp drop-off observed in the \hst-COS spectrum around 3000\AA.  We discuss the implications for the models proposed to explain the sharp drop-off in the UV continuum of \mrk.
For the disk leakage model \citep{2013ApJ...764...15V, 2016ApJ...825...42V}, the sharp drop-off
at the optical-to-NUV bands is due to an increase of extinction at
shorter wavelengths by the intervening absorbers that cover the
optical-to-UV continuum emission region. The flat continuum spectrum
at FUV is due to the emission leaked out of holes in the
spherically distributed absorbers ($\sim$5\% of the sky area).
The leaked emission should still carry the variability signal from the accretion disk, in addition to the scattering dust.  The UV PSD measurement from this paper has constrained the scattering to be smaller than $\sim$10 light days.

The simple BBH model of \citet{2015ApJ...809..117Y} predicts UV variability broadly consistent with the observational results presented in this paper, since the UV emission is from an accretion disk in this model. However, the predicted amount of ionization photons differ significantly depending on detailed model assumptions \citep{2015ApJ...809..117Y, 2016ApJ...829....4L}.
  %Further investigations are needed to re-investigate or expand
    %  the simple BBH model  \citep{2015ApJ...809..117Y} and compare against all available observations.
In addition, the BBH model may suggest a quasi-periodical variation of the UV emission around $\sim$1.2\,yr for \mrk; however, this frequency is outside of the range that can be probed by the current data set.
The PSD of the UV light curve of Mrk\,231 is consistent with a single power-law with no obvious peaks from quasi-periodical variations.
\citet{2014ApJ...788..123L} proposed that the sharp drop-off in the continuum is caused by a special dust extinction curve, which will completely block the FUV and a portion of UV emission from the accretion disk, and thus the model places the UV emission at large scales, such as the starburst contribution \citep{2014ApJ...788..123L}, or FUV photons scattered out of larger scale media \citep{2016ApJ...829....4L}.
The  \emph{Swift} detection of UV variability and the UV PSD shape show that the UV continuum is consistent with emission from the accretion disk and thus incompatible with the special dust extinction model that completely blocks the FUV/UV emission from the disk.
The nature of UV emission of Mrk\,231 is still a mystery, and more observations are needed to better understand this prototype broad absorption line quasar.  For example, in light of the analysis of this paper, measuring the PSD to even higher frequencies will put stronger constraints on the size and origin of the UV emission of \mrk.

We acknowledge the financial support from the NSF grant AST-1413056.
This work is also partially supported by the National Key Program
   for Science and Technology Research and Development (Grant No.\
   2016YFA0400704), the Strategic Priority Program of the Chinese
  Academy of Sciences (Grant No.\ XDB 23040100), and the National
  Natural Science Foundation of China under grant Nos.\ 11690024 and 11390372.
  FS thanks Ian McHardy for useful discussion.
We thank Kayhan Gultekin for helpful discussion.

\begin{table*}
  \centering
  
  \caption{Counts rates of Mrk\,231 core in the X-ray (0.3--8~keV) band from \emph{Swift} XRT observations and AB    magnitude of Mrk\,231 core in uvm2 filter form \emph{Swift} UVOT observations. Table~\ref{tab:obs} is published in its entirety in the electronic edition of the journal. A portion is shown here for guidance regarding its form and content.
\label{tab:obs}}
  \begin{tabular}{cccccc}
  \hline\hline
    Obsid   &   Obs.\ Start Time & XRT          & X-ray (0.3--8\,keV)     & UVM2         & UVM2 \\
        &   JD$-$2457414.53310    & Exposure (s) & $10^{-3}$ (counts/s)      & Exposure (s) & AB Mag \\
   \hline

00032530003 & 0 & 642.23 & $ 10.53 \pm 5.04 $ & 712.16 & $ 17.33 \pm 0.03 $ \\
00032530004 & 2.0560 & 807.67 & $ 8.52 \pm 3.97 $ & 974.68 & $ 17.32 \pm 0.03 $ \\
00032530005 & 4.5230 & 170.18 & $ -0.37 \pm 5.90 $ & 181.62 & $ 17.33 \pm 0.06 $ \\
00032530006 & 6.0496 & 810.47 & $ 13.38 \pm 4.78 $ & 992.62 & ... \\
00032530007 & 8.7092 & 889.76 & $ 14.54 \pm 4.66 $ & 945.25 & $ 17.35 \pm 0.03 $ \\
00032530008 & 10.5587 & 789.24 & $ 6.34 \pm 3.52 $ & 1045.85 & $ 17.32 \pm 0.03 $ \\
00032530010 & 14.6152 & 901.12 & $ 9.92 \pm 3.94 $ & 966.21 & $ 17.31 \pm 0.03 $ \\
00032530011 & 16.6239 & 639.18 & $ 5.99 \pm 4.06 $ & 789.94 & $ 17.31 \pm 0.03 $ \\
00032530012 & 18.8773 & 483.62 & $ 8.15 \pm 5.31 $ & 584.97 & ... \\
00032530013 & 20.8732 & 924.37 & $ 19.17 \pm 5.20 $ & 990.01 & $ 17.32 \pm 0.03 $ \\
00032530014 & 22.6711 & 877.29 & $ 14.66 \pm 4.74 $ & 931.68 & $ 17.31 \pm 0.03 $ \\

   \hline
   \end{tabular}
\end{table*}

%\begin{longtable}{ccc}
%
%    \hline\hline
%Band   &   ${r_{max}}$   &   Lags(day)    \\
%\hline
%SX (0.3--2.0keV)  &  0.91 & $0 \pm 0.79$ \\
%HX (2.0--8.0keV) &  0.94 &  $0 \pm 0.63$ \\
%
%\hline
%    \multicolumn{3}{c}{} \\
%     \caption{{\color{red} The first column is the band for which correlation was measured relative to uvm2 band.
% The second column is the value of maximum correlation coefficient. The third column is the value of
%centroid lag and its $1\sigma$ error.}}
%  \label{tab3}
%\end{longtable}

\end{document}